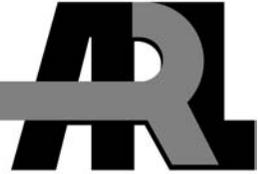

# Force on an Asymmetric Capacitor

by Thomas B. Bahder and Chris Fazi

When a high voltage (~30 kV) is applied to a capacitor whose electrodes have different physical dimensions, the capacitor experiences a net force toward the smaller electrode (Biefeld-Brown effect). We have verified this effect by building four capacitors of different shapes. The effect may have applications to vehicle propulsion and dielectric pumps. We review the history of this effect briefly through the history of patents by Thomas Townsend Brown. At present, the physical basis for the Biefeld-Brown effect is not understood. The order of magnitude of the net force on the asymmetric capacitor is estimated assuming two different mechanisms of charge conduction between its electrodes: ballistic ionic wind and ionic drift. The calculations indicate that ionic wind is at least three orders of magnitude too small to explain the magnitude of the observed force on the capacitor. The ionic drift transport assumption leads to the correct order of magnitude for the force, however, it is difficult to see how ionic drift enters into the theory. Finally, we present a detailed thermodynamic treatment of the net force on an asymmetric capacitor. In the future, to understand this effect, a detailed theoretical model must be constructed that takes into account plasma effects: ionization of gas (or air) in the high electric field region, charge transport, and resulting dynamic forces on the electrodes. The next series of experiments should determine whether the effect occurs in vacuum, and a careful study should be carried out to determine the dependence of the observed force on gas pressure, gas species and applied voltage.

**NOTICES**

**Disclaimers**

The findings in this report are not to be construed as an official Department of the Army position unless so designated by other authorized documents.

Citation of manufacturer's or trade names does not constitute an official endorsement or approval of the use thereof.

Destroy this report when it is no longer needed. Do not return it to the originator.

# Force on an Asymmetric Capacitor

Thomas B. Bahder and Chris Fazi
Sensors and Electron Devices Directorate

# Contents

# Figures

# Acknowledgments

One of the authors (T.B.B.) thanks W. C. McCorkle, Director of Aviation and Missile Command, for the suggestion to look at the physics responsible for the net force on an asymmetric capacitor. The authors want to thank Jean-Louis Naudin (JLN Labs) for his permission to reproduce the letter of Thomas Townsend Brown in Appendix B. One of the authors (T.B.B.) is grateful for personal correspondence with Jean-Louis Naudin (JLN Labs).

# 1. Introduction

Recently, there is a great deal of interest in the Biefeld-Brown effect, i.e., when a high voltage (~30 kV) is applied to the electrodes of an asymmetric capacitor, a net force is observed on the capacitor. By asymmetric, we mean that the physical dimensions of the two electrodes are different, i.e., one electrode is large and the other small. According to the classical Biefeld-Brown effect (see Brown's original 1960, 1962, and 1965 patents cited in Appendix A, and a partial reproduction in section 2), the largest force on the capacitor is in a direction from the negative (larger) electrode toward the positive (smaller) electrode. Today, there are numerous demonstrations of this effect on the Internet in devices called "lifters," which show that the force on the capacitor exceeds its weight [1]. In fact, these experiments indicate that there is a force on the capacitor independent of polarity of applied voltage. In the future, the Biefeld-Brown effect may have application to aircraft or vehicle propulsion, with no moving parts. At the present time, there is no accepted detailed theory to explain this effect, and hence the potential of this effect for applications is unknown. The authors are aware of only two reports [2] and theoretical papers that address such issues [3, 4].

In section 2, we describe the history of the Biefeld-Brown effect. The effect of a net force on an asymmetric capacitor is so surprising that we carried out preliminary simple experiments at the U.S. Army Research Laboratory (ARL) to verify that the effect is real. The results of these experiments are described in section 3. Section 4 contains estimates of the force on the capacitor for the case of ballistic ionic wind and drift of carriers across the capacitor's gap between electrodes. In section 5, we present a detailed thermodynamic treatment of the force on an asymmetric capacitor, assuming that a non-linear dielectric fluid fills the region between capacitor electrodes. Section 6 is a summary and recommendation for future experimental and theoretical work.

# 2. Biefeld-Brown Effect

During the 1920s, Thomas Townsend Brown was experimenting with an x-ray tube known as a "Coolidge tube," which was invented in 1913 by the American physical chemist William D. Coolidge. Brown found that the Coolidge tube exhibited a net force (a thrust) when it was turned on. He believed that he had discovered a new principle of electromagnetism and gravity. Brown applied for a British patent on April 15, 1927, which was issued on November 15, 1928 as Patent No. 300,311, entitled, "Method of Producing Force or Motion." The patent and its figures clearly describe Brown's early work on forces on asymmetric capacitors, although the electromagnetic concepts are mixed with gravitational concepts (Figure 1).

> This invention relates to a method of controlling gravitation and for deriving power therefrom, and to a method of producing linear force or motion. The method is fundamentally electrical.

Figure 1. Excerpt from Thomas Townsend Brown British Patent No. 300,311 entitled "Method of Producing Force or Motion," issued on November 15, 1928.

The discovery of the Biefeld-Brown effect is generally credited to Thomas Townsend Brown. However, it is also named in honor of Brown's mentor, Dr. Paul Alfred Biefeld, a professor of physics and astronomy at Denison University in Granville, Ohio, where Brown was a laboratory assistant in electronics in the Department of Physics. During the 1920s, Biefeld and Brown together experimented on capacitors.

In order to find a technical description of the Biefeld-Brown effect, we performed a search of the standard article literature and found no references to this effect. It is prudent to ask whether this effect is real or rumor. On the other hand, the Internet is full of discussions and references to this effect, including citations of patents issued [1], see also Appendix A. In fact, patents seem to be the only official publications that describe this effect.

On July 3, 1957, Brown filed another patent entitled "Electrokinetic Apparatus," and was issued a U.S. Patent No. 2949550 on August 16, 1960. The effect in this patent is described more lucidly than his previous patent No. 300,311, of November 15, 1928. In this 1960 patent, entitled "Electrokinetic Apparatus," Brown makes no reference to gravitational effects (Figure 2).

> This invention was disclosed and described in my application Serial No. 293,465, filed June 13, 1952, which application has become abandoned. However, reference may be made to this application for the purpose of completing the disclosure set forth below.
>
> The invention utilizes a heretofore unknown electrokinetic phenomenon which I have discovered; namely, that when a pair of electrodes of appropriate form are held in a certain fixed spaced relation to each other and immersed in a dielectric medium and then oppositely charged to an appropriate degree, a force is produced tending to move the pair of electrodes through the medium. The invention is concerned primarily with certain apparatus for utilizing such phenomenon in various manners to be described.

Figure 2. Excerpt from Thomas Townsend Brown U.S. Patent No. 2949550 entitled "Electrokinetic Apparatus," issued on August 16, 1960.

The claims, as well as the drawings in this patent, clearly show that Brown had conceived that the force developed on an asymmetrical capacitor could be used for vehicle propulsion. His drawings in this patent are strikingly similar to some of the capacitors designs on the Internet today. In this 1960 patent, entitled "Electrokinetic Apparatus," Brown gives the clearest explanation of the physics of the Biefeld-Brown effect. Brown makes several important statements, including:

- the greatest force on the capacitor is created when the small electrode is positive
- the effect occurs in a dielectric medium (air)
- the effect can be used for vehicle propulsion or as a pump of dielectric fluid
- Brown's understanding of the effect, in terms of ionic motion
- the detailed physics of the effect is not understood

In the following, we reproduce Brown's first two figures and partial text explaining the effect (Figures 3 and 4).

I have discovered that when apparatus of the character just described is immersed in a dielectric medium, as for example, the ordinary air of the atmosphere, there is produced a force tending to move the entire assembly through the medium, and this force is applied in such direction as to tend to move the body 20 toward the leading electrode 21. This force produces relative motion between the apparatus and the surrounding fluid dielectric. Thus, if the apparatus is held in a fixed position, the dielectric medium is caused to move past the apparatus and to this extent the apparatus may be considered as analogous to a pump or fan. Conversely, if the apparatus is free to move, the relative motion between the medium and the apparatus results in a forward motion of the apparatus, and it is thus seen that the apparatus is a self-propulsive device.

While the phenomenon just described has been observed and its existence confirmed by repeated experiment, the principles involved are not completely understood. It has been determined that the greatest forces are developed when the leading electrode is made positive with respect to the body 20, and it is accordingly thought that in the immediate vicinity of the electrode 21 where the potential gradient is very high, free electrons are stripped off of the atoms and molecules of the surrounding medium. These electrons migrate to the positive electrode 21 where they are collected. This removal of free electrons leaves the respective atoms and molecules positively charged and such charged atoms and molecules are accordingly repelled from the positive electrode 21 and attracted toward the negative electrode 20. The paths of movement of these positively charged particles appear to be of the nature represented by the lines 27 in Figure 2.

It appears that upon reaching or closely approaching the surface of the body 20, the positively charged atoms and molecules have their positive charges neutralized by the capture of electrons from the body 20 and in many cases, it may be that excess electrons are captured whereby to give such atoms and molecules a negative charge so that they are actually repelled from the body 20.

It will be appreciated that the mass of each of the individual electrons is approximately one two-thousandths the mass of the hydrogen atom and is accordingly negligible as compared with the mass of the atoms and molecules of the medium from which they are taken. The principal forces involved therefore are the forces involved in moving the charged atoms and molecules from the region of the positive electrode 21 to and beyond the negatively charged body 20. The force so exerted by the system on those atoms and molecules not only produces a flow of the medium relative to the apparatus, but, of course, results in a like force on the system tending to move the entire system in the opposite direction; that is, to the left as viewed in Figure 1 of the drawing.

The above suggested explanation of the mode of operation of the device is supported by observation of the fact that the dimensions and potentials utilized must be adjusted to produce the required electric field and the resulting propulsive force. Actually I have found that the potential gradient must be below that value required to produce a visible corona since corona is objectionable inasmuch as it represents losses through the radiation of heat, light and molecular charges in the medium.

My experiments have indicated that the electrode 21 may be of small diameter for the lower voltage ranges, i.e. below 125 kv. while above this voltage, rod or hollow pipe electrodes are preferred. These large electrodes are preferred for the higher voltages since sharp points or edges are eliminated which at these elevated potentials would produce losses thus diminishing the thrust. For example, electrodes to be operated at potentials below 125 kv. may be made from small gauge wire only large enough to provide the required mechanical rigidity while electrodes to be operated at potentials above 125 kv. may be hollow pipes or rods having a diameter of ¼ to ½ inch.

In Figure 3, I have illustrated the manner in which a plurality of assemblies, such as are shown in Figure 1, may be interconnected for joint operation. As may be seen from Figure 3, a plurality of such assemblies are placed in spaced side-by-side relation. They may be held fixed in such spaced relation through the use of a plurality of tie rods 28 and interposed spacers (not shown) placed between adjacent plates 20. The assembly of plates 20 may be electrically interconnected by a bus bar or similar conductor 29 to which the negative lead 25 is connected. In a similar way, the plurality of positive leading electrodes 21 may be held in appropriately spaced relation to each other by fastening their ends to pairs of bus bars 30 and 31, to the latter of which the positive lead 26 is connected. The assembly of leading electrodes 21 may be held in spaced relation to the assembly of body members 20 by an appropriate arrangement of the supports 22.

In Figure 4, I have illustrated diagrammatically an arrangement of parts for producing a reversible action; that is, permitting the direction of the propulsive force to be reversed. The apparatus is similar to that shown in Figure 1, differing therefrom in utilizing a pair of leading electrodes 21f and 21r spaced by means of spacers 22 from the front and rear edges 23f and 23r of the body member 20 in a manner similar to that described with reference to the supports 22 in Figure 1. The source 24 of high voltage electrical potential has its negative terminal connected to the body 20 as by means of the aforementioned conductor 25. The positive terminal is connected as by means of the conductor 26 to the blade 27 of a single-pole, double-throw switch, serving in one position to connect the conductor 26 to a conductor 26f which is in turn connected to the forward electrode 21f and arranged in its opposite position to connect the conductor 26 to a conductor 26r which is in turn connected to the reverse electrode 21r.

It will be seen that with the switch 27 in the position shown in Figure 4, the apparatus will operate in the manner described in connection with Figure 1, causing the assembly to move to the left as viewed in Figure 4. By throwing the switch 27 to the opposite position, the direction of the forces produced are reversed and the device moves to the right as viewed in Figure 4.

In Figure 5, I have illustrated the principles of the invention as embodied in a simple form of mobile vehicle. This device includes a body member 50 which is preferably of the form of a circular disc somewhat thicker in its center than at its edges. The disc 50 constitutes one of the electrodes and is the equivalent of the body member 20 referred to in connection with Figure 1. A leading electrode 51 in the form of a wire or similar small diameter conductor is supported from the body 50 by a plurality of insulating supports 52 in uniform spaced parallel relation to a leading edge portion 53 of the body 50. A skirt or similar fairing 54 may be carried by the body 50 to round out the entire structure so as to provide a device which is substantially circular in plan. A source of high voltage electrical potential 55 is provided with its negative terminal connected as indicated at 56 to the body 50 and its positive terminal connected as indicated at 57 to the leading electrode 51.

The device operates in the same manner as the apparatus shown in Figure 1 to produce a force tending to move the entire assembly through the surrounding medium to the left as viewed in Figure 5 of the drawing.

Referring now to Figure 6, there is depicted an illustrative embodiment of this invention in which a pair of mobile vehicles, such as depicted in Figure 5, are shown suspended from the terminals of arm 40, which arm is supported at its midpoint by a vertical column 41. High voltage source 55 is shown connected through wires

Figure 3. Excerpt from Thomas Townsend Brown U.S. Patent No. 2949550 entitled "Electrokinetic Apparatus," issued on August 16, 1960.

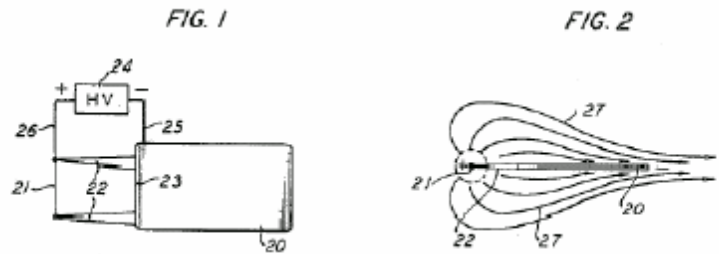

Figure 4. Figure excerpt from Thomas Townsend Brown U.S. Patent No. 2949550 entitled "Electrokinetic Apparatus," issued on August 16, 1960.

Soon after Brown's 1957 filing for the patent previously mentioned, on May 12, 1958, A.H. Bahnson Jr. filed for an improved patent entitled "Electrical Thrust Producing Device," which was granted a U.S. Patent No. 2958790 on November 1, 1960.

On July 3, 1957, Brown filed another patent (granted on January 23, 1962, as U.S. Patent No. 3018394) for an "Electrokinetic Transducer." This patent deals with the inverse effect, i.e., when a dielectric medium is made to move between high voltage electrodes, there is a change in the voltage on the electrodes. (This is reminiscent of Faraday's law of induction.) Quoting from the 1962 patent by Thomas Townsend Brown (Figure 5).

> This invention utilizes heretofore unknown electrokinetic phenomenon which I have discovered, namely that when pairs of electrodes of appropriate form are held in a certain fixed spacial relationship to each other and immersed in a dielectric medium and then oppositely charged to an appropriate degree, a force is produced tending to move the surrounding dielectric with respect to the pair of electrodes. I have also discovered that if the dielectric medium is moved relative to the pairs of electrodes by an external mechanical force, a variation in the potential of the electrodes results which variation corresponds to the variations in the applied mechanical force.
> Accordingly, it is an object of this invention to provide a method and apparatus for converting the energy of an electrical potential directly into a mechanical force suitable for causing relative motion between a structure and the surrounding medium.

Figure 5. Excerpt from Thomas Townsend Brown U.S. Patent No. 3018394 entitled "Electrokinetic Transducer," issued on January 23, 1962.

Until this time, the net force on an asymmetric capacitor was reported as occurring when the capacitor was in a dielectric medium. On May 9, 1958, Brown filed for another patent (improving upon his previous work) entitled "Electrokinetic Apparatus." The patent was issued on June 1, 1965 as Patent No. 3,187,206. The significance of this new patent is that it describes the existence of a net force on the asymmetric capacitor as occurring even in vacuum. Brown states that, "The propelling force however is not reduced to zero when all environmental bodies are removed beyond the apparent effective range of the electric field." Here is a quote from the patent (Figure 6).

> 3,187,206
> ELECTROKINETIC APPARATUS
> Thomas Townsend Brown, Walkertown, N.C., assignor, by mesne assignments, to Electrokinetics, Inc., a corporation of Pennsylvania
> Filed May 9, 1958, Ser. No. 734,342
> 23 Claims. (Cl. 310—5)
>
> This invention relates to an electrical device for producing thrust by the direct operation of electrical fields.
> I have discovered that a shaped electrical field may be employed to propel a device relative to its surroundings in a manner which is both novel and useful. Mechanical forces are created which move the device continuously in one direction while the masses making up the environment move in the opposite direction.
> When the device is operated in a dielectric fluid medium, such as air, the forces of reaction appear to be present in that medium as well as on all solid material bodies making up the physical environment.
> In a vacuum, the reaction forces appear on the solid environmental bodies, such as the walls of the vacuum chamber. The propelling force however is not reduced to zero when all environmental bodies are removed beyond the apparent effective range of the electrical field.
> By attaching a pair of electrodes to opposite ends of a dielectric member and connecting a source of high electrostatic potential to these electrodes, a force is produced in the direction of one electrode provided that electrode is of such configuration to cause the lines-of-force to converge steeply upon the other electrode. The force, therefore, is in a direction from the region of high flux density toward the region of low flux density, generally in the direction through the axis of the electrodes. The thrust produced by such a device is present if the electrostatic field gradient between the two electrodes is non-linear. This non-linearity of gradient may result from a difference in the configuration of the electrodes, from the electrical potential and/or polarity of adjacent bodies, from the shape of the dielectric member, from a gradient in the density, electric conductivity, electric permittivity and manetic permeability of the dielectric member or a combination of these factors.

Figure 6. Excerpt from Thomas Townsend Brown Patent No. 3,187,206, entitled, "Electrokinetic Apparatus," issued on June 1, 1965.

In this patent, Brown reports that the asymmetric capacitor does show a net force, even in vacuum. However, at present, there is little experimental evidence, except for two reports [2], which do not explain the origin of the observed force and two theoretical papers [3, 4]. If the Biefeld-Brown effect is to be understood on a firm basis, it is imperative to determine whether the effect occurs in vacuum. Enclosed in Appendix B, is my email correspondence with J. Naudin, where Naudin quotes from a letter by Thomas Townsend Brown, who discusses the effect in vacuum.

The main question to be answered is: what is the physical mechanism that is responsible for the net force on an asymmetric capacitor? The answer to this question may depend on whether the asymmetric capacitor is in a polarizable medium (such as air), or in vacuum. However, to date, the physical mechanism is unknown, and until it is understood, it will be impossible to determine its potential for practical applications.

## 3. Preliminary Experiments at ARL

The Biefeld-Brown effect is reported in many places on the Internet; however, as mentioned above, only two papers exist [3, 4]. Therefore, we decided to verify that the effect was real. C. Fazi (ARL) and T. Bahder (ARL) have fabricated three simple asymmetric capacitors, using the designs reported on the Internet [1]. In all three cases, we have verified that a net force is exerted on the capacitors when a high DC voltage is applied to the electrodes. The three asymmetric capacitors that we tested had different geometries, but they all had the common feature that one electrode was thin and the other very wide (asymmetric dimensions). Also, a suspended wire, representing a capacitor with the second electrode at infinity, showed lift.

Our first model was made by Tom Bahder, and was triangular in shape, which is a typical construction reported on the Internet (Figure 7). One electrode is made from thin 38-gauge (0.005-mil) wire, and the other electrode is made from ordinary Aluminum foil. The capacitor is ~20 cm on a side, the foil sides are 20 cm × 4 cm, and the distance of the top of the foil to the thin wire electrode is 3 cm. The foil and wire are supported by a Balsa wood frame, so that the whole capacitor is very light, ~5 g. Initially, we made the Balsa wood frame too heavy (capacitor weight ~7 grams), and later we cut away much of the frame to lighten the construction to ~5 g. We found that in order to demonstrate the lifting effect, the capacitor must be made of minimum weight. (Typical weights reported on the Internet for the design in Figure 7 are 2.3 g–4 g.)

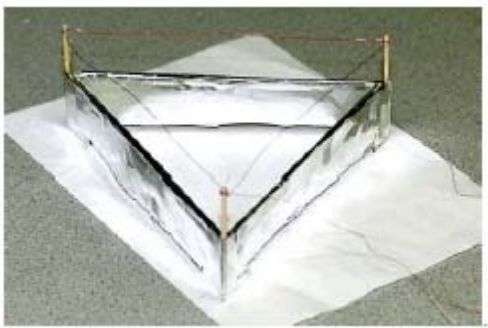

Figure 7. Our first attempt at making an asymmetric capacitor (a "lifter"), according to the specifications given by J. Naudin on Internet Web site <http://jnaudin.free.fr/>.

When ~37 kV was applied to the capacitor in Figure 7, the current was ~1.5 mA. The capacitor lifted off its resting surface. However, this capacitor was not a vigorous flier, as reported by others on the Internet. One problem that occurred was arcing from the thin wire electrode to the foil. The thin wire electrode was too close to the foil. We have found that arcing reduces the force developed on the capacitor. Also, compared to other constructions, ours was too heavy, 5 g. We found that a ground plane beneath the capacitor is not essential for the lifting force to exceed the capacitor's weight.

Consequently, we decided to make a second version of an asymmetric capacitor, using a Styrofoam lunch box and plastic drinking straws from the ARL cafeteria (Figure 8). The capacitor had a square geometry, 18 cm × 20 cm. The distance of the thin wire (38 gauge) to the foil was adjustable, and we found that making a 6-cm gap resulted in little arcing. When 30 kV was applied, the capacitor drew ~1.5 mA, and hovered vigorously above the floor.

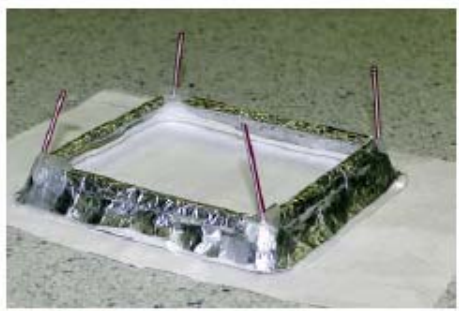

Figure 8. The second attempt at making a lighter asymmetric capacitor.

A question occurred: is the toroidal (closed circular) geometry of the capacitor electrodes essential to the lifting effect that we have observed? Consequently, Tom Bahder made a flat-shaped, or wing-shaped, capacitor as shown in Figure 9. This capacitor was made from two (red) plastic coffee stirrers and a (clear) plastic drinking straw to support the Aluminum foil. The significance of the clear plastic straw was that the foil could be wrapped over it, thereby avoiding sharp foil edges that would lead to corona discharge or arcing. The dimensions of the foil on this capacitor were 20 cm × 4 cm, as shown in Figure 9. The distance between the thin wire electrode (38 gauge wire) and edge of the foil was 6.3 cm. This capacitor showed a net force on it when ~30 kV was applied, drawing ~500 mA. The force on this capacitor greatly exceeded its weight, so much so that it would vigorously fly into the air when the voltage was increased from zero. Therefore, we have concluded that the closed geometry of the electrodes is not a factor in the net force on an asymmetric capacitor. Furthermore, the force on the capacitor always appeared in the direction toward the small electrode—independent of the orientation of the capacitor with respect to the plane of the Earth's surface. The significance of this observation is that the force has nothing to do with the gravitational field of the Earth and nothing to do with the electric potential of the Earth's atmosphere. (There are numerous claims on the Internet that asymmetric capacitors are antigravity devices, or devices that demonstrate that there is an interaction of gravity with electric phenomena, called.)

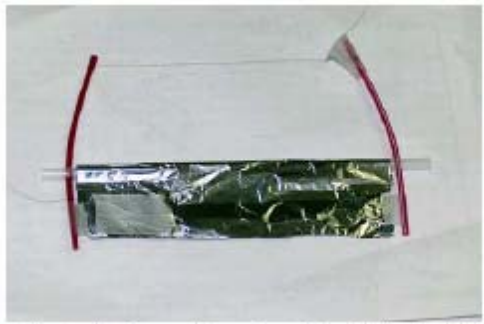

Figure 9. Flat-shaped (or wing-shaped) asymmetric capacitor used to test whether closed electrode geometry is needed.

The thin wire electrode must be at a sufficient distance away from the foil so that arcing does not occur from the thin wire electrode to the foil at the operating voltage. In fact, in our first model, shown in Figure 7, the 3-cm gap from top of the foil to thin wire electrode was not sufficiently large, and significant arcing occurred. We have found that when arcing occurs, there is little net force on the capacitor. An essential part of the design of the capacitor is that the edges of the foil, nearest to the thin wire, must be rounded (over the supporting Balsa wood, or plastic straw, frame) to prevent arcing or corona discharge at sharp foil edges (which are closest to the thin wire). The capacitor in Figure 7 showed improved lift when rounded foil was put over the foil electrode closest to the thin wire, thereby smoothing-over the sharp foil edges. Physically, this means that the radius of curvature of the foil nearest to the small wire electrode was made larger, creating a greater asymmetry in radii of curvature of the two electrodes.

When operated in air, the asymmetric capacitors exhibit a net force toward the smaller conductor, and in all three capacitors, we found that this force is independent of the DC voltage polarity.

The detailed shape of the capacitor seems immaterial, as long as there is a large asymmetry between the characteristic size of the two electrodes. The simplest capacitor configuration consists of a suspended thin wire from the hot electrode of the high-voltage power supply, as shown in Figure 10. To observe the wire movement, a small piece of transparent tape was attached at the lower end of the thin wire. A suspended thin wire (~12 in length) also showed force with ~35 kV and 1-mA current (Figure 9). From a vertical position, the wire lifted, as shown in Figure 11by as much as 30°, once the high voltage approached 35 kV. The usual air breakdown hissing sound of the other capacitors was heard when current reached ~1 mA. Actually, the wire did not remain suspended, but oscillated back and forth approximately 60° from vertical, and the hissing pitch followed the oscillation period with amplitude and frequency changes. Without the piece of tape at the end, the wire did not lift as much and the sound was considerably weaker. The piece of tape seems to increase the capacitance and or the air ionization. This suspended wire configuration can be viewed also as a capacitor surrounded by the ground system located several feet away (metallic benches, floor and ceiling). As in the other capacitor experiments, it also did not exhibit a polarity dependence.

When the asymmetric capacitors have an applied DC voltage, and they are producing a net force in air, they all emit a peculiar hissing sound with pitch varying with the applied voltage. This sound is similar to static on a television or radioset when it is not tuned to a good channel. We believe that this sound may be a clue to the mechanism responsible for the net force.

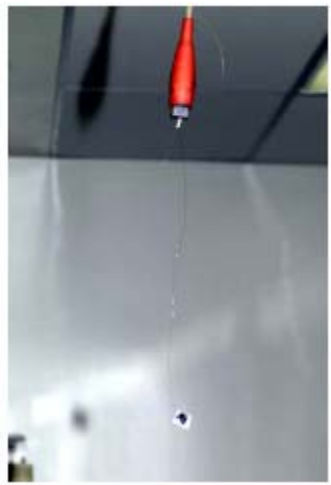

Figure 10. The capacitor consisting of a single wire. No bias applied.

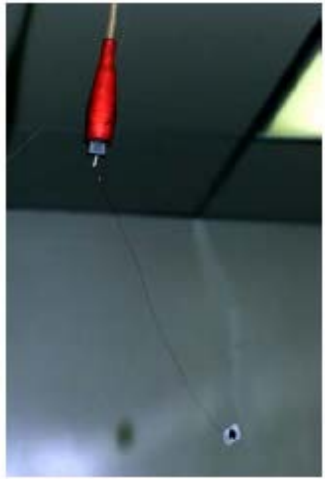

Figure 11. The wire capacitor showing displacement from the vertical. 35 kV applied.

## 4. Previously Proposed Explanations for the Biefeld-Brown Force

There are two proposed explanations for the Biefeld-Brown force. Both of these have been discussed on the Internet in various places. The first proposed scheme is that there exists an ionic wind in the high field region between the capacitor electrodes, and that this ionic wind causes the electrodes to move as a result of the momentum recoil. This scheme, described in Section 4.1, leads to a force that is incorrect by at least three orders of magnitude compared to what is observed. (This scheme also assumes ballistic transport of charges in the atmosphere between electrodes of the capacitor, and it is known that instead drift current exists.)

In section 4.2, we present the second scheme, which assumes that a drift current exists between the capacitor plates. This scheme is basically a scaling argument, and not a detailed treatment of the force. In this scheme, the order of magnitude of the force on an asymmetric capacitor is correct, however, this scheme is only a scaling theory. Finally, in section 5, we present our thermodynamic treatment of the force on an asymmetric capacitor.

### 4.1 Ionic Wind: Force Too Small

The most common explanation for the net force on an asymmetric capacitor invokes ionic wind. Under a high-voltage DC bias, ions are thought to be accelerated by the high potential difference between electrodes, and the recoil force is observed on an asymmetric capacitor. A simple upper limit on the ion wind force shows that the ion wind effect is at least three orders of magnitude too small. Consider a capacitor that operates at voltage V. Charged particles of mass $m$, having charge $q$, such as electrons or (heavy) ions, are accelerated to a velocity $v$, having a kinetic energy

$$\frac{1}{2}mv^2 = qV. \tag{1}$$

The force exerted on an asymmetric capacitor is given by the rate of change of momentum

$$F = mv\frac{I}{q}, \tag{2}$$

where $I$ is the current flowing through the capacitor gap, and we assume that all the ionic momentum, $mv$, is transferred to the capacitor when the charged particles leave an electrode. Also, we assume that none of this momentum is captured at the other electrode. This is a gross over-estimation of the force due to ionic effects, so equation (2) is an upper limit to the ionic force.

Solving equation (1) for the velocity, and using it in equation (2) gives the upper limit on the force due to ionic wind

$$F = \left(\frac{2mV}{q}\right)^{\frac{1}{2}} I. \tag{3}$$

When the force $F$ is equal to the weight of an object, $Mg$, where $g$ is the acceleration due to gravity, the force will lift a mass

$$M = \left(\frac{2mV}{q}\right)^{\frac{1}{2}} \frac{I}{g}. \tag{4}$$

If we assume that electrons are the charged particles responsible for force of the ionic wind, then we must use mass $m = 9.1 \times 10^{-31}$ kg. Substituting typical experimental numbers into equation (4), I find that the ionic wind can lift a mass

$$M = \left(\frac{(2)(9.1 \times 10^{-31}\text{kg})(40 \times 10^3 \text{ Volt})}{1.6 \times 10^{-19} \text{ C}}\right)^{\frac{1}{2}} \frac{1.0 \times 10^{-3} \text{ A}}{10\,\frac{\text{m}}{\text{s}^2}} = 6.8 \times 10^{-5} \text{ gram.} \tag{5}$$

The typical weight of an asymmetric capacitor is on the order of 5 g, so this force is too small by 5 orders of magnitude.

Another possibility is that heavy ions (from the air or stripped off the wire) are responsible for the ionic wind. As the heaviest ions around, assume that Cu is being stripped from the wire. Using Cu for the ions, the mass of the ions is 63.55 $m_p$, where 63.55 is the atomic mass of Cu and $m_p$ is the mass of a proton. The weight that could be lifted with Cu ionic wind is then (upper limit):

$$M = \left( \frac{(2)(63.55)(1.67 \times 10^{-27}\,\text{kg})(40 \times 10^{3}\,\text{Volts})}{1.6 \times 10^{-19}\,\text{C}} \right)^{\frac{1}{2}} \frac{1.0 \times 10^{-3}\,\text{A}}{10\,\frac{\text{m}}{s^{2}}} = 0.002\,\text{gram}. \qquad (6)$$

Again, this value is three orders of magnitude too small to account for lifting a capacitor with a mass of 3–5 g. Therefore, the ionic wind contribution is too small, by at least three orders of magnitude, to account for the observed force on an asymmetric capacitor.

While the force of the ionic wind computed above is too small to explain the experiments in air, it should be noted that this effect will operate in vacuum, and may contribute to the overall force on a capacitor.

**4.2 The Ion Drift Picture: Scaling Theory of Force**

In the previous section, we computed an upper limit to the force on a capacitor due to ionic wind effects. Ionic wind is a ballistic flow of charges from one electrode to the other. Clearly the force due to ionic wind is at least three orders of magnitude too small to account for the observed force on an asymmetric capacitor (in air). There is another type of classical transport: drift of charge carriers in an electric field. In the case of drift, the carriers do not have ballistic trajectories, instead they experience collisions on their paths between electrodes. However, due to the presence of an electric field, the carriers have a net motion toward the opposite electrode. This type of transport picture is more accurate (than ballistic ionic wind) for a capacitor whose gap contains air. Drift transport is used by Evgenij Barsoukov to explain the net force on an asymmetric capacitor [5].

The general picture of the physics is that the positive and negative electrodes of the capacitor are charged and that these charges experience different forces because the electric field surrounding the capacitor is nonuniform (Figure 12. The electric field surrounding the capacitor is created by the potential applied to the capacitor electrodes and partial ionization of air into positive ions and electrons. These charge carriers experience drift and diffusion in the resulting electric field. The battery supplies the energy that is dissipated by transport of carriers in the electric field. The electric field is particularly complicated because it is the result of a steady state: the interplay between the dynamics of ionization of the air in the high-field region surrounding the electrodes and charge transport (drift and diffusion of positive and negative carriers) in the resulting electric field.

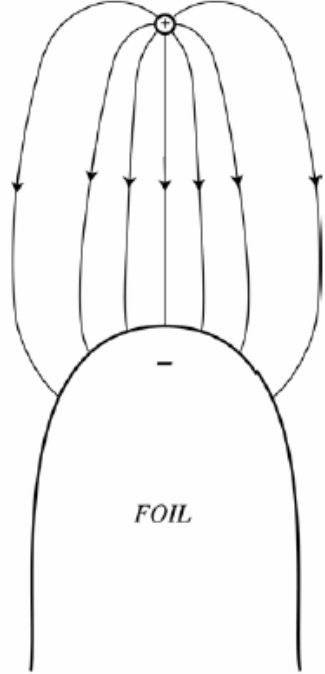

Figure 12. Schematic diagram of the side view of electric field for the asymmetric capacitor in Figure 9.

If the capacitor is surrounded by vacuum (rather than a dielectric, such as ions on air), the net force *F* on the asymmetric capacitor can be computed by the sum of two surface integrals, one over the surface of the positive electrode and one over the surface of the negative electrode [6]:

$$F = \frac{1}{2}\varepsilon_0 \left( \int_{S_+} E^2 \, \mathbf{n} \, dS + \int_{S_-} E^2 \, \mathbf{n} \, dS \right), \qquad (7)$$

where $\varepsilon_0$ is the permittivity of vacuum, *E* is the electric field normal to the conducting electrodes, $S_+$ and $S_-$ are the positive and negative electrode surfaces of the capacitor and **n** is the outward normal to $S_+$ and $S_-$. The integrals in equation (7) are done over closed surfaces $S_+$ and $S_-$. As stated above, the complexity of the calculation is contained in computing the electric field *E*. In section 5, we give an expression for the net force on the capacitor assuming that it is surrounded by a dielectric, such as air.

The electric field around the small wire electrode is much stronger than the field around the foil (see Figures 9 and 12). In our experiments, there is a big difference in the radii of curvature of the two capacitor electrodes: the thin wire electrode has a radius $r_1$ = 0.0025 inch, and the edge of the foil has a radius of curvature of $r_2$ = 0.125 inch. This difference in curvature leads to an electric field with a strong gradient. The ratios of electric fields at the thin wire electrode to that at the rounded edge of the foil is inversely proportional to the square of the radii of curvatures: $E_1/E_2 = (r_1/r_2)^2 \sim 2500$. However, the applied voltage is on the order of 30 kV, over a gap of 6 cm, so an electric field of magnitude 2500 ×30 kV/6 cm ~ $1 \times 10^7 \frac{V}{cm}$ would not be supported in air. It is clear that screening of the electric field is occurring due to the dielectric effects of charged air ions and electrons, as well as polarized air atoms. When a positive high voltage is applied to the thin wire electrode of the asymmetric capacitor, ionization of air atoms, such as Nitrogen, probably occurs first near the thin wire electrode. The ionization of Nitrogen atoms leads to free electrons and ions near the small electrode. The electron mobility is significantly larger for electrons than for Nitrogen ions. This can be expected because the current density $J = \sigma E = n\, e\, v$ where $\sigma = n\, e^2\, \tau / m$ is the electrical conductivity, $n$ is charge density, $\tau$ is the scattering time, and the mean drift velocity $v = \mu E$. So the mobility behaves as $\mu = e\, \tau / m$. Because electrons are three orders of magnitude more massive than ions, it is expected that they are correspondingly more mobile. Experimentally, it is found that the electron mobility in air at atmospheric pressure and electric field $E = 10^4$ Volt/cm is approximately [7]

$$\mu_e = 620 \frac{cm^2}{Volt \cdot sec}. \tag{8}$$

The mobility of $N_2$ ions in air is [8]

$$\mu_{N_2} = 2.5 \frac{cm^2}{Volt \cdot sec}. \tag{9}$$

Therefore, the physical picture is that in the high field region the electrons, with their high mobility, are swept out by the electric field, toward the thin wire electrode leading to screening of the field. The massive (probably positive) ions are less mobile and are left behind in a plasma surrounding the thin wire electrode.

A scaling argument can be made as follows: The lower foil conductor feels a force $F$ of magnitude

$$F = Q \frac{V}{\ell}, \tag{10}$$

where $Q$ is the charge on the foil electrode, $V$ is the voltage between the capacitor conductors, and $l$ is the length of the gap between thin wire electrode and foil. The charge $Q$ and voltage $V$ are quantities that are actually present when screening is taking place. The negative charge on the foil, $-Q$, can be approximated in terms of the measured current, $I \sim$ 1mA, by saying that all the carriers are swept out in a time $t$ :

$$I = \frac{Q}{t} = Q \frac{v}{\ell}, \tag{11}$$

where $t$ is the time for carriers to move across the capacitor gap, $l$, if they are travelling at an average drift velocity, $v$. Eliminating the charge $Q$ from equations (10) and (11), leads to an expression for the net force on the capacitor

$$F = I \frac{V}{v}. \tag{12}$$

In equation (12), the current $I$ is a measured quantity, the voltage $V$ is on the order of 30 kV, and the drift velocity for electrons is [7]

$$v_e = 6.2 \times 10^6 \frac{\text{cm}}{\text{sec}}. \tag{13}$$

Alternatively, the electron drift velocity, $v_e$, can be expressed in terms of the mobility, $\mu_e$, given in equation (8), and electric field, $E$. The net force on the asymmetric capacitor is then given by

$$F = I \frac{V}{\mu E} = I \frac{l}{\mu} \tag{14}$$

where we again used $E = V/\ell$. Using the value of electron mobility in equation (8), the net force becomes

$$F = I \frac{l}{\mu} = \frac{(10^{-3}\text{ A})(0.04\text{ m})}{\left(620 \frac{\text{cm}^2}{\text{Volt} \cdot \text{sec}}\right)\left(10^{-2} \frac{\text{m}}{\text{cm}}\right)^2} = 6.4 \times 10^{-4}\text{ N}. \tag{15}$$

The force in equation (15) could lift a mass $M$

$$M \frac{F}{g} = \frac{6.4 \times 10^{-4}\text{ N}}{10 \frac{\text{m}}{\text{s}^2}} = 0.064 \text{ gram}. \tag{16}$$

The typical asymmetric capacitor has a mass that is two orders or magnitude greater. Consequently, drift of electrons cannot explain the observed force on the capacitor.

An alternative to using the value of electron mobility is to use the smaller value of ionic mobility. This will lead to a larger force because the force in equation (14) is inversely proportional to the mobility.

$$F = I \frac{l}{\mu} = \frac{(10^{-3}\text{ A})(0.04\text{ m})}{\left(2.5 \frac{\text{cm}^2}{\text{Volt} \cdot \text{sec}}\right)\left(10^{-2} \frac{\text{m}}{\text{cm}}\right)^2} = 0.16 \text{ N}. \tag{17}$$

The force in equation (17), due to the drift of Nitrogen ions, could lift a mass $M$:

$$M = \frac{F}{g} = \frac{0.16\text{ N}}{10 \frac{\text{m}}{\text{s}^2}} = 16 \text{ gram}. \tag{18}$$

The force on the capacitor, given in equation (18), is within a factor of 3, assuming a capacitor of mass 5 g.

As alternative derivation of the scaling equation (14), consider the asymmetric capacitor as being essentially an electric dipole of magnitude,

$$|\mathbf{p}| = p = Ql, \tag{19}$$

where $Q$ is the charge on one plate and $l$ is the average effective separation between plates. When a high voltage is applied to the asymmetric capacitor (assume positive voltage on the thin wire and negative on the foil), the high electric field around the thin wire ionizes the atoms of the air. There is comparatively little ionization near the foil due to the lower magnitude electric field near the foil. The ionized atoms around the foil form a plasma, consisting of charged electrons and positively charged ions. The force on the capacitor must scale like

$$\mathbf{F} = \nabla(\mathbf{p} \cdot \mathbf{E}) \tag{20}$$

where $E$ is the electric field. The gradient operates on the electric field, producing a magnitude $dE/dx \sim E/\ell$. Using this value in equation (20), together with the size of the dipole in equation (19), leads to a force on the capacitor

$$F = Q\frac{V}{\ell} \sim \frac{I\ell}{v} \cdot \frac{V}{\ell} = I\frac{V}{v}, \tag{21}$$

which is identical to equation (12).

From the scaling derivations that were presented, it is clear that electron drift current leads to a force on the capacitor that is too small. Using the value of mobility appropriate for (nitrogen) ions leads to a force whose order of magnitude is in agreement with experiment.

Note that the force, given by equation (14), scales inversely with the mobility $\mu$. If the ions are responsible for providing the required small mobility, then the picture is that the ions are like a low-mobility molasses, which provides a large spacecharge to attract the negatively charged foil electrode. As soon as the foil electrode moves toward the positive ion cloud, another positive ionic cloud is set up around the thin electrode, using the energy from the voltage source. In this way, the dipole (asymmetric capacitor) moves in the nonuniform electric field that it has created. Physically, this is a compelling picture; however, much work must be done (experimentally and theoretically) to fill in important details to determine if this picture has any merit.

## 5. Thermodynamic Analysis of the Biefeld-Brown Force

In this section, we present our hypothesis that the Biefeld-Brown force, generated on an asymmetric capacitor, can be described by the thermodynamics of a fluid dielectric in an external electric field produced by charged conductors. The (partially ionized) air between capacitor electrodes is the fluid dielectric. Although the air is partially ionized, we assume that this fluid dielectric is close to neutral on the macroscopic scale. The charged conductors are the asymmetric electrodes of the capacitor. The battery provides the charge on the electrodes and the energy to sustain the electric field in the air (dielectric) surrounding the capacitor electrodes.

The total system is composed of three parts: the partially ionized air dielectric, the metal electrodes of the capacitor and the battery (voltage source) and connecting wires, and the electromagnetic field. The battery is simply a large reservoir of charge. The total momentum (including the electromagnetic field) of this system must be constant [9]:

$$\mathbf{P}_{dielectric} + \mathbf{P}_{electrodes} + \mathbf{P}_{field} = \text{constant}, \tag{22}$$

where **P**<sub>dielectric</sub> is the momentum of the fluid dielectric (air in the capacitor gap and surrounding region), **P**<sub>electrodes</sub> is the momentum of the metallic electrodes, wire and battery, and **P**<sub>field</sub> is the momentum of the electromagnetic field. Taking the time derivative of equation (22), the forces must sum to zero

$$\mathbf{F}_{dielectric} + \mathbf{F}_{electrodes} + \frac{d\mathbf{P}_{field}}{dt} = 0. \qquad (23)$$

As far as the electric field is concerned, its total momentum changes little during the operation of the capacitor, because the field is in a steady state; energy is supplied by the battery (charge reservoir). So we set the rate of change of field momentum to zero, giving a relation between the force on the electrodes and the dielectric:

$$\mathbf{F}_{electrodes} = -\mathbf{F}_{dielectric}. \qquad (24)$$

A lengthy derivation based on thermodynamic arguments leads to an expression for the stress tensor, $\sigma_{ik}$, for a dielectric medium in an electric field [6, 10, 11,

$$\sigma_{ik} = \left[\tilde{F} - \rho\left(\frac{\partial \tilde{F}}{\partial \rho}\right)_{T,\mathbf{E}}\right]\delta_{ik} + E_i D_k, \qquad (25)$$

where the free energy $\tilde{F}$ is a function of the fluid density, $\rho$, temperature, $T$, and electric field **E**. The differential of the free energy is given by

$$d\tilde{F} = -S\,dT + \zeta\,d\rho - \mathbf{D}\cdot d\mathbf{E}, \qquad (26)$$

where $S$ is the entropy, **D** is the electric induction vector, and $\zeta$ is the chemical potential per unit mass [6]. Equation (25) is valid for any constitutive relation between **D** and **E**. We assume that the air in between the capacitor plates is an isotropic, but nonlinear, polarizable medium, due to the high electric fields between plates. Therefore, we take the relation between **D** and **E** to be

$$\mathbf{D} = \varepsilon(E)\mathbf{E}, \qquad (27)$$

where $\varepsilon(E)$ is a scalar dielectric function that depends on the magnitude of the electric field, $E = |\mathbf{E}|$, the temperature, $T$, and the density of the fluid, $\rho$. We have suppressed the dependence of $\varepsilon$ on $T$ and $\rho$ for brevity. The dielectric function $\varepsilon(E)$ depends on position through the variables $T$ and $\rho$ and because the medium (air) between capacitor plates is assumed to be non-uniform. Inserting equation (27) into equation (26), we integrate the free energy along a path from $E = 0$ to some finite value of $E$ obtaining

$$\tilde{F}(\rho,T,\mathbf{E}) = \tilde{F}_o(\rho,T) - \frac{1}{2}\varepsilon_{eff}\,E^2 \qquad (28)$$

where $\varepsilon_{eff}$ is an effective (averaged) dielectric constant, given by

$$\varepsilon_{eff} = \frac{1}{E^2}\int_0^{E^2}\varepsilon\!\left(\sqrt{\xi}\right)d\xi \qquad (29)$$

where $\xi$ is a dummy integration variable. The dielectric constant $\varepsilon_{\text{eff}}$ depends on spatial position (because of $\varepsilon$), on $T$, $\rho$, and on electric field magnitude $E$.

The body force per unit volume of the dielectric, $f_i$, is given by the divergence of the stress tensor,

$$f_i = \frac{\partial \sigma_{ik}}{\partial x_k}, \tag{30}$$

where there is an implied sum over the repeated index $k$. Performing the indicated differentiations in equation (30), we obtain an expression for the body force [6,10,11]

$$\mathbf{f} = -\nabla P_o(\rho, T) + \frac{1}{2}\nabla\left[E^2 \rho \left(\frac{\partial \varepsilon_{\text{eff}}}{\partial \rho}\right)_{T,\mathbf{E}}\right] - \frac{1}{2}E^2 \nabla \varepsilon_{\text{eff}} + \frac{1}{2}(\varepsilon - \varepsilon_{\text{eff}})\nabla E^2 + \rho_{\text{ext}}\mathbf{E}, \tag{31}$$

where the external charge density is give by div $\mathbf{D} = \rho_{\text{ext}}$. This charge density is the overall external charge density in the dielectric, which may have been supplied by the battery, electrodes, and the surrounding air. In equation (31), the pressure $P_o(\rho, T)$ is that which would be present in the absence of the electric field. In the case of a linear medium, the dielectric function $\varepsilon$ is independent of field $E$, and $\varepsilon_{\text{eff}} = \varepsilon$, which reduces to the result derived by Landau and Lifshitz (see their equations (15.12) in reference[6]).

The total force on the fluid dielectric, $\mathbf{F}_{\text{dielectric}}$, is given by the volume integral of $\mathbf{f}$ over the volume of the dielectric, $\Omega$:

$$\mathbf{F}_{\text{dielectric}} = \int_{\Omega} \mathbf{f}\, dV \tag{32}$$

The volume $\Omega$ is the whole volume outside the metal electrodes of the capacitor. According to equation (24), the net force on the capacitor, $\mathbf{F}_{\text{electrodes}}$, is the negative of the total force on the dielectric:

$$\mathbf{F}_{\text{electrodes}} = \int_{\Omega}\left\{\frac{1}{2}E^2 \nabla \varepsilon + \frac{1}{2}\nabla\left[(\varepsilon_{\text{eff}} - \varepsilon)E^2\right] - \frac{1}{2}\nabla\left[E^2 \rho\left(\frac{\partial \varepsilon_{\text{eff}}}{\partial \rho}\right)_{T,\mathbf{E}}\right] - \rho_{\text{ext}}\mathbf{E}\right\} dV, \tag{33}$$

where we have dropped the term containing the gradient in the pressure, assuming that it is negligible. Equation (33) gives the net force on capacitor plates for the case where the fluid dielectric is nonlinear, having the response given in equation (27). In equation (33), both $\varepsilon$ and $\varepsilon_{\text{eff}}$ are functions of the electric field. Note that the first three terms of the integrand depend on the square of the electric field, which is in agreement with the fact that the observed force direction is independent of the polarity of the applied bias.

There are four terms in the force. The first term is proportional to the gradient of the dielectric constant, $\nabla \varepsilon$. We expect that the dielectric constant has a large variation in between regions of low and high electric field, such as near the smaller electrode. We expect that there is a strong nonlinear dielectric response due to ionization of the air. The resulting free charges can move large distances, leading to a highly nonlinear response at high electric fields. Therefore , it is possible that this first term in the integrand in equation (33) has the dominate contribution. We expect this term to contribute to a force that points toward the smaller electrode (as observed experimentally), and we expect that this contribution is nearly independent of polarity of applied bias.

The second term in the force equation (33) is proportional to the gradient of the product of the square of the electric field and the difference in dielectric constants. The difference in the dielectric constants, $\varepsilon_{\text{eff}} - \varepsilon$, can be expanded in a Taylor series in $E$

$$\varepsilon_{\text{eff}} - \varepsilon = -\frac{1}{3}\varepsilon'(0)E - \frac{1}{4}\varepsilon''(0)E^2 + \dots, \tag{34}$$

where

$$\varepsilon'(0) = \left(\frac{\partial \varepsilon}{\partial E}\right)_{T,\rho,\mathbf{E}=\mathbf{0}}, \text{ and}$$

$$\varepsilon''(0) = \left(\frac{\partial^2 \varepsilon}{\partial E^2}\right)_{T,\rho,\mathbf{E}=\mathbf{0}}. \tag{35}$$

The gradient of the square of the electric field always points toward the smaller electrode, independent of the polarity of bias applied to the capacitor. We do not know the sign of the dielectric constants $\varepsilon'(0)$ and $\varepsilon''(0)$. If the air has dielectric properties described by $\varepsilon'(0) < 0$ and $\varepsilon''(0) < 0$, then this term would contribute to a force toward the smaller electrode (which would be in agreement with experiment). Alternatively, the term $\frac{1}{2}\nabla\left[(\varepsilon_{\text{eff}} - \varepsilon)E^2\right]$ may have the wrong sign but may be small. This must be determined experimentally by studying the dielectric properties of air or other gas.

The third term in the force equation (33) is difficult to evaluate. It may well be negligible, especially compared to the first term (assuming highly nonlinear dielectric response at high fields). Alternatively, if the air behaves as a nearly linear dielectric medium, then $\varepsilon_{\text{eff}} \sim \varepsilon$, and the dielectric constant of a gas is typically proportional to its density, $\varepsilon = \alpha\, \varepsilon_o\, \rho$, where $\varepsilon_o$ is the permittivity of free space, and $\alpha$ is a constant. Using these expressions in equation (33) for $\varepsilon$ yields the force on the capacitor electrodes for the case of a linear dielectric fluid:

$$\left(F_{\text{electrodes}}\right)_{\text{Linear Medium}} = \int_\Omega \left\{-\frac{1}{2}\varepsilon\nabla E^2 - \rho_{\text{ext}}\mathbf{E}\right\} dV. \tag{36}$$

For a linear medium, the first term in equation (35) contributes to a force pointing in a direction that is opposite to the gradient of the square of the electric field, i.e., it points toward the larger electrode (opposite to the experimentally observed force). In order to obtain a net force from equation (36) that is oriented toward the smaller electrode, the second term in equation (36) would have to dominate, i.e., the net force on the capacitor would be due to external charge effects. The magnitude of the external charges (from battery and surrounding air) on the dielectric fluid must be determined experimentally.

If the space between the capacitor plates is filled with a vacuum instead of dielectric, equation (33) reduces to a force given by

$$\left(\mathbf{F}_{\text{electrodes}}\right)_{\text{Vacuum}} = -\int_{\Omega} \rho_{\text{ext}} \mathbf{E}\, dV. \tag{37}$$

where $\rho_{\text{ext}} = 0$ for vacuum, leading to zero force on the capacitor.

The thermodynamic theory presented here provides a general expression in equation (33) for the net force on a capacitor in terms of the macroscopic electric field **E**. This electric field in equation (33) must be determined by a microscopic calculation, taking into account the ionization of gas between capacitor plates, and details of charge transport.

In summary, at the present time, the relative magnitudes of the fours terms in the force expression given in equation (33) are unknown. The magnitudes of these terms must be determined by constructing a set of experiments designed to determine the field-dependent dielectric properties of the fluid (given by $\varepsilon$) surrounding the asymmetric capacitor electrodes. These experiments will permit us to verify if the thermodynamic theory presented here can explain the magnitude and sign of the observed force.

## 6. Summary and Suggested Future Work

We have presented a brief history of the Biefeld-Brown effect: a net force is observed on an asymmetric capacitor when a high voltage bias is applied. The physical mechanism responsible for this effect is unknown. In section 4, we have presented estimates of the force on the capacitor due to the effect of an ionic wind and due to charge drift between capacitor electrodes. The force due to ionic wind is at least three orders of magnitude too small. The force due to charge drift is plausible, however, the estimates are only scaling estimates, not a microscopic model.

In section 5, we have presented a detailed thermodynamic theory of the net force on a capacitor that is immersed in a nonlinear dielectric fluid, such as air in a high electric field. The main result for the net force on the capacitor is given in equation (33). The thermodynamic theory requires knowledge of the dielectric properties of the fluid surrounding the capacitor plates. It is not possible to estimate the various contributions to the force until we have detailed knowledge about the high-field dielectric properties of the fluid.

More experimental and theoretical work is needed to gain an understanding of the Biefeld-Brown effect. As discussed, the most pressing question is whether the Biefeld-Brown effect occurs in vacuum. It seems that Brown may have tested the effect in vacuum, but not reported it (Appendix B). More recently, there is some preliminary work that tested the effect in vacuum, and claimed that there is some small effect—smaller than the force observed in air; see the second report cited in reference [2]. Further work must be done to understand the effect in detail. A set of experiments must be performed in vacuum, and at various gas pressures, to determine the force versus voltage and current. A careful study must be made of the force as a function of gas species and gas pressure. In order to test the thermodynamic theory presented here, the dielectric properties of the gas must be carefully measured. Obtaining such data will be a big step toward developing a theoretical explanation of the effect. On the theoretical side, a microscopic model of the capacitor (for a given geometry) must be constructed, taking into account the complex physics of ionization of air (or other gas) in the presence of high electric fields. Only by understanding the Biefeld-Brown effect in detail can its potential for applications be evaluated.

# Appendix A.  Short Patent History Dealing With Asymmetric Capacitors

Townsend Brown, T.  "A Method of and an Apparatus or Machine for Producing Force or Motion."  GB Patent 300311 issued on November 15, 1928[6].

Townsend Brown, T.  "Electrokinetic Apparatus."  U.S. Patent 2949550 issued on August 16, 1960.

Bahnson, A. H. Jr.  "Electrical thrust producing device."  U.S. Patent 2958790 issued on November 1, 1960.

Townsend Brown, T.  "Electrokinetic Transducer."  U.S. Patent 3018394 issued on January 23, 1962.

Townsend Brown, T.  "Electrokinetic Apparatus."  U.S. Patent 3187206 issued on June 1, 1965.

Bahnson, A.H. Jr.  "Electrical thrust producing device."  U.S. Patent 3227901 issued on January 4, 1966.

Cambell, J. W. (NASA).  "Apparatus for Generating Thrust Using a Two Dimensional, Asymmetrical Capacitor Module."  U.S. Patent US2002012221, issued January 31, 2002.

Cambell, J. W. (NASA).  "Aparatus for Generating Thrust Using a Two Dimensional Asymmetric Capacitor Module."  U.S. Patent 6411493 issued on June 25, 2002.

# Appendix B.  Force on Asymmetric Capacitor in Vacuum

Enclosed below is a copy of my email correspondence with Jean-Louis Naudin (JLN Labs) [1], who hosts a Web site on "Lifters."  In this correspondence, Naudin quotes a letter, purportedly signed by T. Townsend Brown, in which Brown discusses the question of whether an asymmetric capacitor has a net force on it in vacuum under high voltage.

T. Townsend Brown's letter, as provided by J. Naudin:

*Dear ....,*

*You have asked several question which I shall try to answer. The experiments in vacuum were conducted at "Societe Nationale de Construction Aeronautique" in Paris in 1955-56, in the Bahnson Laboratories, Winston-Salem, North Carolina in 1957-58 and at the "General Electric Space Center" at King of Prussia, Penna, in 1959.*

*Laboratory notes were made, but these notes were never published and are not availible to me now.  The results were varied, depending upon the purpose of the experiment.  We were aware that the thrust on the electrode structures were caused largely by ambiant ion momentum transfer when the experiments were conducted in air.  Many of the tests, therefore, were directed to the exploration of this component of the total thrust. In the case of the G.E.  test, cesium ions were seeded into the environment and the additional thrust due to seeding was observed.*

*In the Paris test miniature saucer type airfoils were operated in a vaccum exceeding 10-6mm Hg.Bursts of thrust (towards the positive) were observed every time there was a vaccum spark within the large bell jar.- These vacuum sparks represented momentary ionization, principally of the metal ions in the electrode material.  The DC potential used ranged from 70kV to 220kV.*

*Condensers of various types, air dielectric and barium titanate were assembled on a rotary support to eliminate the electrostatic effect of chamber walls and observations were made of the rate of rotation.Intense acceleration was always observed during the vacuum spark (which, incidentally, illuminated the entire interior of the vacuum chamber).  Barium Titanate dielectrique always exceeded air dielectric in total thrust.  The results which were most significant from the -standpoint of the Biefeld-Brown effect was that thrust continued, even when there was no vacuum spark, causing the rotor to accelerate in the negative to positive direction to the point where voltage had to be reduced or the experiment discontinued because of the danger that the rotor would fly apart.*

*In short, it appears there is strong evidence that Biefeld-Brown effect does exist in the negative to positive direction in a vacuum of at least 10-6 Torr.  The residual thrust is several orders of magnitude larger than the remaining ambient ionization can account for.Going further in your letter of January 28th, the condenser "Gravitor" as described in my British patent, only showed a loss of weight when vertically oriented so that the negative-to-postive thrust was upward.  In other words, the thrust tended to "lift" the gravitor.  Maximum thrust observed in 1928 for one*

*gravitor weighing approximately 10 kilograms was 100 kilodynes at 150kV DC. These gravitors were very heavy, many of them made with a molded dielectric of lead monoxide and beeswax and encased in bakelite. None of these units ever "floated" in the air.*

*There were two methods of testing, either as a pendulum, in which the angle of rise against gravity was measured and charted against the applied voltage, or, as a rotor 4ft. in diameter, on which four "gravitors" were mounted on the periphery. This 4 ft. wheel was tested in air and also under transformer oil.The total thust or torque remained virtually the same in both instances, seeming to prove that aero-ionization was not wholly responsible for the thrust observed.Voltage used on the experiments under oil could be increased to about 300kV DC and the thrust appeared to be linear with voltage.*

*In subsequent years, from 1930 to 1955, critical experiments were performed at the Naval Research Laboratory, Washington, DC.; the Randall-Morgan Laboratory of Physics, University of Penna., Philadelphia; at a field station in Zanesvill, Ohio, and two field stations in Southern California, of the torque was measured continuously day and night for many years. Large magnitude variations were consistenly observed under carefully controlled conditions of constant voltage, temperature, under oil, in magnetic and electrostatic shields, not only underground but at various elevations. These variations, recorded automatically on tape, were statistically processed and several significant facts were revealed.*

*There were pronounced correlations with mean solar time, sideral time and lunar hour angle. This seemed to prove beyond a doubt that the thrust of "gravitors" varied with time in a way that related to solar and lunar tides and sideral correlation of unknown origin. These automatic records, acquired in so many different locations over such a long period of time, appear to indicate that the electrogravitic coupling is subject to an extraterrestrial factor, possibly related to the universal gravitational potential or some other (as yet) unidentified cosmic variable.In response to additional questions, a reply of T.T. Brown, dated April, 1973, stated :"The apparatus which lifted itself and floated in the air, which was described by Mr Kitselman, was not a massive dielectric as described in the English patent.Mr Kitselman witnessed an experiment utilising a 15" circular, dome-shaped aluminum electrode, wired and energized as in the attached sketch. When the high voltage was applied, this device, althrough tethered by wires from the high voltage equipment, did rise in the air, lifting not only its own weight but also a small balance weight which was attached to it on the uderside. It is true that this apparatus would exert a force upward of 110% of its weight.*

*The above experiment was an improvement on the experiment performed in Paris in 1955 and 1956 on disc air foils. The Paris experiments were the same as those shown to Admiral Radford in Pearl Harbor in 1950.*

*These experiments were explained by scientific community as due entirely to "ion-momentum transfer", or "electric wind". It was predicted categorically by many "would-be" authorities that such an apparatus would not operate in vaccum. The Navy rejected the research proposal (for further research) for this reason. The experiments performed in Paris several years later, proved that ion wind was not entirely responsible for the observed motion and proved quite conclusively that the apparatus would indeed operate in high vacuum.*

*Later these effects were confirmed in a laboratory at Winston-Salem, N.C., especially constructed for this purpose. Again continuous force was observed when the ionization in the*

*medium surrounding the apparatus was virtually nil.In reviewing my letter of April 5th, I notice, in the drawing which I attached, that I specified the power supply to be 50kV. Actually, I should have indicated that it was 50 to 250kV DC for the reason that the experiments were conducted throughout that entire range.*

*The higher the voltage, the greater was the force observed. It appeared that, in these rough tests, that the increase in force was approximately linear with voltage. In vaccum the same test was carried on with a canopy electrode approximately 6" in diameter, with substantial force being displayed at 150 kV DC. I have a short trip of movie film showing this motion within the vacuum chamber as the potential is applied."*

*Kindest personal regards,*

*Sincerely,*

*T.Townsend Brown*

| REPORT DOCUMENTATION PAGE | | | Form Approved<br>OMB No. 0704-0188 |
|---|---|---|---|
| Public reporting burden for this collection of information is estimated to average 1 hour per response, including the time for reviewing instructions, searching existing data sources, gathering and maintaining the data needed, and completing and reviewing the collection information. Send comments regarding this burden estimate or any other aspect of this collection of information, including suggestions for reducing the burden, to Department of Defense, Washington Headquarters Services, Directorate for Information Operations and Reports (0704-0188), 1215 Jefferson Davis Highway, Suite 1204, Arlington, VA 22202-4302. Respondents should be aware that notwithstanding any other provision of law, no person shall be subject to any penalty for failing to comply with a collection of information if it does not display a currently valid OMB control number.<br>**PLEASE DO NOT RETURN YOUR FORM TO THE ABOVE ADDRESS.** | | | |
| **1. REPORT DATE** (DD-MM-YYYY)<br>March 2003 | **2. REPORT TYPE**<br>Final | | **3. DATES COVERED** (From - To)<br>August 2002 to December 2002 |
| **4. TITLE AND SUBTITLE**<br>Force on an Asymmetric Capacitor | | | **5a. CONTRACT NUMBER** |
| | | | **5b. GRANT NUMBER** |
| | | | **5c. PROGRAM ELEMENT NUMBER**<br>62705A |
| **6. AUTHOR(S)**<br>Thomas B. Bahder and Christian Fazi | | | **5d. PROJECT NUMBER**<br>3NE6BC |
| | | | **5e. TASK NUMBER** |
| | | | **5f. WORK UNIT NUMBER** |
| **7. PERFORMING ORGANIZATION NAME(S) AND ADDRESS(ES)**<br>U.S. Army Research Laboratory<br>Attn: AMSRL-SE-EE<br>2800 Powder Mill Road<br>Adelphi, MD 20783-1197 | | | **8. PERFORMING ORGANIZATION REPORT NUMBER**<br>ARL-TR- |
| **9. SPONSORING/MONITORING AGENCY NAME(S) AND ADDRESS(ES)**<br>U.S. Army Research Laboratory<br>2800 Powder Mill Road<br>Adelphi, MD 20783-1197 | | | **10. SPONSOR/MONITOR'S ACRONYM(S)** |
| | | | **11. SPONSOR/MONITOR'S REPORT NUMBER(S)** |
| **12. DISTRIBUTION/AVAILABILITY STATEMENT**<br>Approved for public release; distribution unlimited | | | |
| **13. SUPPLEMENTARY NOTES**<br>AMS Code 622705.H94<br>DA Project AH94 | | | |
| **14. ABSTRACT**<br>When a high voltage (~30 kV) is applied to a capacitor whose electrodes have different physical dimensions, the capacitor experiences a net force toward the smaller electrode (Biefeld-Brown effect). We have verified this effect by building four capacitors of different shapes. The effect may have applications to vehicle propulsion and dielectric pumps. We review the history of this effect briefly through the history of patents by Thomas Townsend Brown. At present, the physical basis for the Biefeld-Brown effect is not understood. The order of magnitude of the net force on the asymmetric capacitor is estimated assuming two different mechanisms of charge conduction between its electrodes: ballistic ionic wind and ionic drift. The calculations indicate that ionic wind is at least three orders of magnitude too small to explain the magnitude of the observed force on the capacitor. The ionic drift transport assumption leads to the correct order of magnitude for the force, however, it is difficult to see how ionic drift enters into the theory. Finally, we present a detailed thermodynamic treatment of the net force on an asymmetric capacitor. In the future, to understand this effect, a detailed theoretical model must be constructed that takes into account plasma effects: ionization of gas (or air) in the high electric field region, charge transport, and resulting dynamic forces on the electrodes. The next series of experiments should determine whether the effect occurs in vacuum, and a careful study should be carried out to determine the dependence of the observed force on gas pressure, gas species and applied voltage. | | | |
| **15. SUBJECT TERMS**<br>Electrostatic propulsion, capacitor, high voltage, dielectric, ion propulsion, Bieheld-Brown effect, thermodynamics, force, electric | | | |
| **16. SECURITY CLASSIFICATION OF:** | | | **17. LIMITATION OF ABSTRACT** | **18. NUMBER OF PAGES** | **19a. NAME OF RESPONSIBLE PERSON**<br>Thomas B. Bahder |
| **a. REPORT**<br>UNCLASSIFIED | **b. ABSTRACT**<br>UNCLASSIFIED | **c. THIS PAGE**<br>UNCLASSIFIED | UL | 34 | **19b. TELEPHONE NUMBER** (Include area code)<br>301-394-2044 |

**Standard Form 298 (Rev. 8/98)**
Prescribed by ANSI Std. Z39.18